\documentclass[reprint, NumberedRefs]{JASA}

\usepackage{placeins}
\usepackage[utf8]{inputenc}
\usepackage[english]{babel}
\usepackage{graphics}
\usepackage{graphicx}
\usepackage{amsmath,amssymb,esint}
\usepackage{multirow}
\usepackage{amsmath}
\usepackage{comment}
\usepackage{upgreek}
\usepackage{booktabs}
\usepackage{lineno}

\allowdisplaybreaks

\newcommand{\changes}[1]{{\color{black}{#1}}}

\begin{document}

\title{Acoustic pressure modulation driven by spatially non-uniform flow}

\author{Fabian Denner}
\email{fabian.denner@polymtl.ca}
\affiliation{Department of Mechanical Engineering, Polytechnique Montr\'eal, Montr\'eal, H3T 1J4, QC, Canada}
\date{\today}

\begin{abstract}
    The recent identification of a modulation of acoustic waves that is driven by spatial velocity gradients, using acoustic black and white hole analogues [Schenke et al., {J.~Acoust.~Soc.~Am.}~154 (2023), 781-791], has shed new light on the complex interplay of acoustic waves and non-uniform flows. According to the virtual acoustic black hole hypothesis, these findings should be applicable to acoustic waves propagating in non-uniform flows of arbitrary velocity. In this study, the propagation of acoustic waves in non-uniform flows is investigated by incorporating a leading-order model of the acoustic pressure modulation into a Lagrangian wave tracking algorithm. Using this numerical method, the acoustic pressure modulation is recovered accurately in non-uniform subsonic flows. This suggests that spatial velocity gradients drive acoustic pressure modulations in any non-uniform flow, which can, as shown here, be readily quantified.
\end{abstract}


\maketitle


\section{Introduction}

Acoustic waves are modulated by the fluid they propagate in, the motion of their  source and the prevailing flow conditions. \citep{Morfey1971,Ostashev2016} \changes{Non-uniform flows are known to act upon acoustic waves in complex ways and directly influence jet noise, \citep{Edgington-Mitchell2019} the sound emitted by rotors, \citep{Carley2020} and the efficiency of acoustic cloaking. \citep{He2019} The modulations imparted on acoustic waves also encode distinct information about the environment in which the wave has propagated and about the motion of the associated acoustic source or scatterers. This information can be utilized in acoustic monitoring and remote sensing \citep{Dowling2015} related, for instance, to the terrestrial environment and atmosphere, \citep{Bradley2008} underwater remote sensing and imaging, \citep{Fitzpatrick2020} and medical applications. \citep{Christensen-Jeffries2020} To this end, isolating and distinguishing the physical mechanisms modulating acoustic waves are critical for a precise interpretation of the information encoded in acoustic signals.}

\begin{figure*}[t]
    \includegraphics[width=0.9\textwidth]{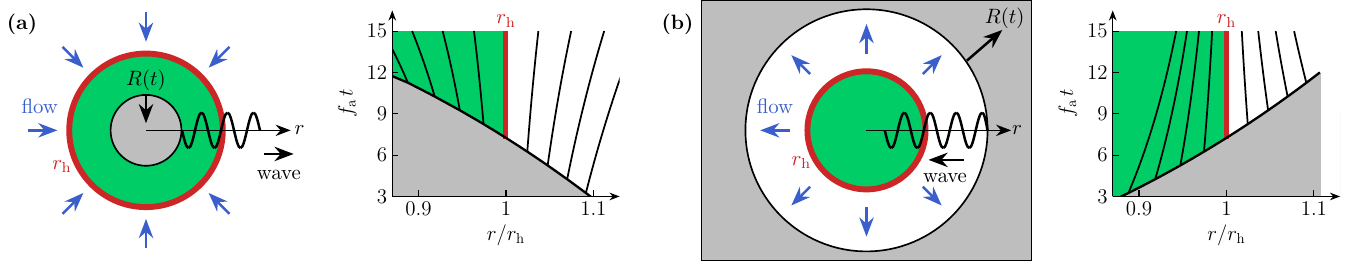}
    \caption{Schematic illustration of (a) a spherical acoustic black hole analogue \changes{and (b) a spherical acoustic white hole analogue,} alongside the corresponding wave characteristics in a space-time diagram. The moving wave-emitting boundary with time-dependent radius $R(t)$ displaces the fluid such that a spatially non-uniform flow with a stationary sonic horizon $r_\mathrm{h}$, shown in red, is formed. The white (green) areas correspond to the subsonic (supersonic) flow regions, the gray area represents the emitter. \changes{The acoustic waves are subject to an accelerating flow in the acoustic black hole, leading to diverging wave characteristics, and to a decelerating flow in the acoustic white hole, leading to converging wave characteristics.}}
    \label{fig:abh}
\end{figure*}

\changes{The analogy between acoustics and relativity offers a powerful tool to gain new insights into both subjects. \citep{Visser1998, Gregory2015, Stone2000, Schenke2022d} To date, this analogy has mainly been adopted in the form of acoustic black and white hole analogues to study the properties of gravitational black holes, \citep{Unruh1981, Mayoral2011} such as Hawking radiation, in laboratory experiments, exploiting that acoustic waves are effectively governed by a (3+1)-dimensional Lorentzian space-time geometry. \citep{Unruh1981, Visser1998} However, the utility of such analogues is not limited to cosmology and quantum physics.}
With the aim of studying nonlinear acoustics, we proposed a spherical acoustic black hole analogue, \citep{Schenke2022d} illustrated in Fig.~\ref{fig:abh}(a), \changes{whereby a spherical acoustic emitter collapses such that a spatially non-uniform flow with a stationary sonic horizon is formed.} The sonic horizon of this acoustic black hole provides a stationary observation point where the propagation speed of acoustic waves vanishes ($c+u=0$), but where the flow has a non-zero acceleration. \changes{This model system, therefore, allows to isolate the influence of an accelerating flow} on acoustic waves. Using this acoustic black hole analogue, together with the corresponding acoustic white hole analogue illustrated in Fig.~\ref{fig:abh}(b), our work identified an amplitude modulation of acoustic waves in accelerating flows, whereby an accelerating (decelerating) flow causes a decreasing (increasing) acoustic pressure at the sonic horizon. \citep{Schenke2023} At the sonic horizon, this pressure modulation is described to leading order, as a function of time $t$ and speed of sound $c$, by \citep{Schenke2023}
\begin{equation}
    \Delta p(t) =  \Delta p_\mathrm{a,\mathrm{h}} \, \frac{1}{1 + \dfrac{a_\mathrm{h} t}{c}}, \label{eq:dpa_corr}
\end{equation}
where $a_\mathrm{h}$ and $\Delta p_\mathrm{a,\mathrm{h}}$ are the convective acceleration and the emitted acoustic pressure at the sonic horizon.

Based on these findings, we postulated the \textit{virtual acoustic black hole hypothesis}, \citep{Schenke2023} stipulating that any flow can be viewed locally as an acoustic black/white hole by applying a Galilean transformation in space-time, \changes{given that the equations governing mass and momentum conservation, as well as the flow acceleration, are invariant under a Galilean transformation. \citep{Wang2022d} Therefore,} any acoustic wave characteristic (i.e.~the trajectories of acoustic particles in space-time) can be assumed to locally form an apparent sonic horizon; a real sonic horizon where $c+u = 0$ is not required. If this virtual acoustic black hole hypothesis holds, which has yet to be demonstrated, the acoustic pressure modulation observed in acoustic black/white hole analogues applies to flows of arbitrary velocity. 

This study scrutinizes the virtual acoustic black hole hypothesis and its promise to quantify the acoustic pressure modulation driven by non-uniform flows of arbitrary velocity. To this end, acoustic waves emitted by spherical emitters with time-varying radius are considered, whereby the boundary motion of the emitter induces a flow that is non-uniform in space and time. \changes{The information associated with the emitted acoustic waves is tracked numerically using a Lagrangian wave tracking algorithm, which is, by default, not able to predict the acoustic pressure modulation associated with non-uniform flows. Therefore, this wave tracking method allows us to test the virtual acoustic black hole hypothesis by introducing a suitable correction for the acoustic pressure.}

\changes{
\section{Pressure modulation in non-uniform flow}
\label{sec:energy}
To elucidate the origin of pressure modulations in a non-uniform flow, we first consider the conservation of energy of acoustic waves in a one-dimensional flow with velocity $u$ and pressure $p$. 
The velocity and pressure are formally decomposed \citep{Visser1998} into background contributions, denoted with subscript $0$, and acoustic perturbations, denoted with subscript $\text{a}$, such that $u = u_0 + u_\text{a}$ and $p=p_0 + p_\text{a}$. Considering small-amplitude acoustic waves, the density $\rho$ and speed of sound $c$ of the fluid are taken to be constant, and the acoustic velocity contribution $u_\text{a} = p_\text{a}/(\rho c)$ is assumed to be negligible ($u_\text{a} \ll u_0$).

Under these assumptions, the acoustic energy of a plane acoustic wave with length $\lambda$ and constant cross-sectional area $A$ is given as
\begin{equation}
E_\lambda = \int_\lambda \frac{p_\text{a}^2}{\rho c^2} \, \text{d}V = \frac{A}{\rho c^2} \int_0^\lambda p_\text{a}^2 \, \text{d}x.\label{eq:acousticenergy}
\end{equation}
For a sinusoidal acoustic wave with pressure amplitude $\Delta p_\text{a}$, the integral of the acoustic pressure becomes
\begin{equation}
    \int_0^\lambda p_\text{a}^2 \, \text{d}x = \int_0^\lambda \Delta p_\text{a}^2 \, \text{sin}^2(kx) \, \text{d}x = \frac{\lambda \, \Delta p_\text{a}^2}{2}. \label{eq:p1_int}
\end{equation}
Inserting Eq.~\eqref{eq:p1_int} into Eq.~\eqref{eq:acousticenergy} yields a first-order approximation for the acoustic energy of a planar sinusoidal acoustic wave,
\begin{equation}
    E_\lambda = \frac{A}{2 \rho c^2}   \, {\lambda \,  \Delta p_\text{a}^2} \propto \lambda \, \Delta p_\text{a}^2.
\end{equation}
By linearizing the background flow over a sufficiently small time interval $\delta t = t_\text{II}-t_\text{I}$, the wavelength follows as \citep{Schenke2023}
\begin{equation}
    \lambda (t_\text{II}) = \lambda(t_\text{I}) \left(1+\dfrac{\partial u_0}{\partial x}  \, \delta t\right).
\end{equation}
Since $ \lambda(t_\text{I}) \, \Delta p_\text{a}(t_\text{I})^2= \lambda(t_\text{II}) \, \Delta p_\text{a}(t_\text{II})^2$, the pressure amplitude is given to leading order as
\begin{equation}
    \Delta p_\text{a}(t_\text{II})^2 = \frac{\Delta p_\text{a}(t_\text{I})^2}{1+\dfrac{\partial u_0}{\partial x} \, \delta t}. \label{eq:p1_modulation_general}
\end{equation}
This suggests that the change in wavelength driven by a spatially non-uniform flow results in a modulation of the acoustic pressure amplitude due to the conservation of acoustic energy and that this change in pressure can be quantified to leading order by Eq.~\eqref{eq:p1_modulation_general}.

Contrary to the planar wave considered above, the cross-sectional area of an acoustic wave emitted by a spherical emitter is not constant as it travels radially outwards. A first-order approximation of the change in cross-sectional area over one wavelength yields \citep{Schenke2023} $\Delta A \propto \lambda$. Consequently, in spherical symmetry the first-order approximation of the conservation of acoustic energy dictates $E_\lambda \propto \lambda^2 \, \Delta p_\text{a}^2$, with $ \lambda(t_\text{I})^2 \, \Delta p_\text{a}(t_\text{I})^2= \lambda(t_\text{II})^2 \, \Delta p_\text{a}(t_\text{II})^2$. Eq.~\eqref{eq:p1_modulation_general}, hence, becomes
\begin{equation}
    \Delta p_\text{a}(t_\text{II}) = \frac{\Delta p_\text{a}(t_\text{I})}{1+\dfrac{\partial u_0}{\partial r} \, \delta t}, \label{eq:p1_modulation_spherical}
\end{equation}
where $r$ denotes the radial coordinate.
Inserting the spatial velocity gradient at the sonic horizon of a spherical acoustic black or white hole analogue, $(\partial u_0/\partial r)_{r_\text{h}} = a_\text{h} / c$, into Eq.~\eqref{eq:p1_modulation_spherical} yields an expression equivalent to Eq.~\eqref{eq:dpa_corr}, 
which is consistent with the virtual acoustic black hole hypothesis. \citep{Schenke2023}
}

\section{Mathematical model}

\changes{The leading-order model for the acoustic pressure modulation in a non-uniform flow presented in Section \ref{sec:energy} is tested by conducting numerical simulations of acoustic waves emitted by a spherical emitter with time-varying radius. The radial motion of this emitter induces a non-uniform flow. For the considered flow in spherical symmetry, the equations governing the conservation of mass and momentum are given as
\begin{eqnarray}
    \frac{\partial \rho}{\partial t} + \frac{1}{r^2} \, \frac{\partial}{\partial r} \left(r^2 \rho u \right) &=& 0 \label{eq:continuity} \\
    \frac{\partial u}{\partial t} + u \frac{\partial u}{\partial r} + \frac{1}{\rho} \frac{\partial p}{\partial r} &=& 0, \label{eq:momentum}
\end{eqnarray}
where $t$ is time and $r$ is the radial coordinate. 
Because in spherical symmetry there is by definition no flow in the azimuthal and polar directions, the flow in the radial direction is irrotational and the velocity can be expressed by the potential $\phi$ as $u =-{\partial \phi}/{\partial r}$.
Integrating the momentum equation, Eq.~(\ref{eq:momentum}), from $r$ to $\infty$ yields the transient Bernoulli equation \citep{Gilmore1952}}
\begin{equation}
    -\frac{\partial \phi}{\partial t} + \frac{u^2}{2} + \frac{p}{\rho} =0,
    \label{eq:momentum_int}
\end{equation}
assuming the potential vanishes at infinity, and the density $\rho$ and speed of sound $c$ are constant.
Rearranging Eq.~\eqref{eq:momentum_int} for the time derivative and multiplying by $r$ yields 
\begin{equation}
    g = r \frac{\partial \phi}{\partial t} = r \left( \frac{p}{\rho} + \frac{u^2}{2} \right). \label{eq:g}
\end{equation}
Following the Kirkwood-Bethe hypothesis, \citep{Kirkwood1942,Denner2023} the quantity $g$ is constant along an outgoing characteristic of a spherical emitter and propagates with speed $c+u$.

The considered spherical emitter with time-varying radius $R(t)$ emits monochromatic acoustic waves with frequency $f_\mathrm{a}$ and amplitude $\Delta p_\mathrm{a}$, with the acoustic pressure at the moving wave-emitting boundary defined as $p_\mathrm{a}(R,t) = \Delta p_\mathrm{a} \, \sin (2\pi f_\mathrm{a} t)$. The radial location along a wave characteristic is given as 
\begin{equation}
    r(t) = R(\tau) + \int_\tau^t [c + u(r,t)] \, \mathrm{d}t, \label{eq:r_t}
\end{equation}
where the retarded time $\tau$ represents the time of origin of the characteristic at the wave-emitting boundary. 
The quantity $g$, Eq.~\eqref{eq:g}, is assumed to be invariant and defined by the boundary conditions applied at the wave-emitting boundary ($r = R$) \changes{at the time of emission ($t=\tau$)} as
\begin{equation}
    g = R(\tau) \left[ \frac{p_\mathrm{a}(R,\tau)}{\rho} + \frac{\dot{R}(\tau)^2}{2} \right]. \label{eq:g_tau}
\end{equation}
Rearranging Eq.~\eqref{eq:g} for $p$ and inserting Eq.~\eqref{eq:g_tau} for $g$, the pressure follows as \citep{Denner2023}
\begin{equation}
    p(r,t) = \frac{R(\tau)}{r(t)} \left[p_\mathrm{a}(R,\tau) + \frac{\rho}{2} \dot{R}(\tau)^2 \right] - \frac{\rho}{2} u(r,t)^2.
    \label{eq:p_rt}
\end{equation}
\changes{Furthermore, the velocity $u$ of the incompressible flow in spherical symmetry follows from the conservation of mass, Eq.~\eqref{eq:continuity}, as 
\begin{equation}
    u(r,t) = \frac{R(t)^2}{r(t)^2} \dot{R}(t). \label{eq:u0_rt}
\end{equation}
Eqs.~\eqref{eq:p_rt} and \eqref{eq:u0_rt} readily satisfy the boundary conditions $p(R,\tau) = p_\mathrm{a}(R,\tau)$ and $u(R,\tau) = \dot{R}(t)$ at the wave-emitting boundary.}

\section{Lagrangian wave tracking}
\label{sec:lagrangian}

\begin{figure}
    \includegraphics[width=0.45\textwidth]{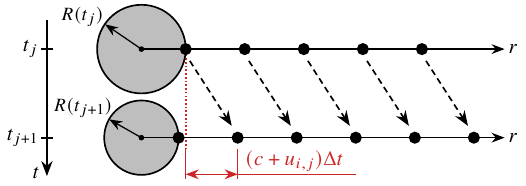}
    \caption{Illustration of the Lagrangian wave tracking method, in which discrete emission nodes are tracked in space and time as they propagate with speed $c+u(r,t)$.}
    \label{fig:tracking}
\end{figure}

The information of the acoustic emissions is tracked numerically using the Lagrangian wave tracking algorithm available in the open-source software library {\tt APECSS}, \citep{Denner2023a} which propagates the information of an acoustic emission in the form of discrete nodes along the outgoing characteristics, \citep{Denner2023} as sketched in Fig.~\ref{fig:tracking}. 
At time instance $t_j$, the radial position $r_{i,j}$ of emission node $i$ is determined numerically based on Eq.~\eqref{eq:r_t} as 
\begin{equation}
    r_i(t_j) \approx r_{i,j} = R(\tau) + \sum_{k=0}^{j-1} \left[c + u_{i,k} \right] \Delta t,\label{eq:r_num}
\end{equation}
using a constant time-step $\Delta t$. For the considered spherical emitter, the pressure and flow velocity at emission node $i$ follow from Eqs.~\eqref{eq:p_rt} and \eqref{eq:u0_rt} as
\begin{align}
    p(r_{i,j},t_j) &\approx p_{i,j} = \frac{R(\tau)}{ r_{i,j}} \, p_{\mathrm{a},i,j} + \frac{\rho}{2} \left[ \frac{R(\tau)}{r_{i,j}} \,  \dot{R}(\tau)^2 - u_{i,j}^2 \right] \label{eq:p_num_corr}\\
    u(r_{i,j},t_j) &\approx u_{i,j} = \frac{R^2(t_j)}{r_{i,j}^2} \dot{R}(t_j),
\end{align}
\changes{where the second term on the right-hand side of Eq.~\eqref{eq:p_num_corr} represents the dynamic pressure contributions of the moving boundary and the flow.}

\changes{At the sonic horizon ($ r= r_\mathrm{h}$) of an acoustic black hole, the flow velocity is $u(r_\mathrm{h},t)=-c$ and, since $c + u(r_\mathrm{h},t) = 0$, the radial location of the acoustic information associated with the sonic horizon remains unchanged, as per Eq.~\eqref{eq:r_num}. Hence, the velocity terms in Eq.~\eqref{eq:p_num_corr} cancel. By default, the pressure at the sonic horizon is, therefore, constant and given as $p(r_\mathrm{h},t) =p_\mathrm{a}(R,\tau)$. This stands in contradiction to the reported acoustic pressure modulation driven by the non-uniform flow at the sonic horizon. \citep{Schenke2023}}

\changes{To account for the acoustic pressure modulation driven by a spatially non-uniform flow, the acoustic pressure $p_{\mathrm{a},i,j}$ in Eq.~\eqref{eq:p_num_corr} is proposed to be corrected by incorporating}  a modulation factor $\mathcal{A}$, such that
\begin{equation}
    p_{\mathrm{a},i,j} =  p_\mathrm{a}(R,\tau) \, \prod_{k=1}^{j} \mathcal{A}_{i,k}. \label{eq:pstar}
\end{equation}
\changes{Assuming the acoustic energy is conserved, the modulation factor $\mathcal{A}_{i,k}$ for the spherical emitters considered here follows from the leading-order model of the pressure modulation in spherical symmetry, Eq.~\eqref{eq:p1_modulation_spherical}, as}
\begin{equation}
    \mathcal{A}_{i,k} = \frac{1}{1 + \dfrac{\partial u_{i,k}}{\partial r} \, \Delta t}, \label{eq:A}
\end{equation}
\changes{where the time interval $\delta t$ is now represented by the numerical time-step $\Delta t$.}

\section{Results}
\label{sec:results}

\changes{Two spherical emitters are considered: a spherical acoustic black hole is used to validate the proposed correction of the Lagrangian wave tracking algorithm, and the virtual acoustic black hole hypothesis is tested using a pulsating spherical emitter that induces a subsonic flow. 
Because only acoustic waves with small pressure amplitude are considered, the pressure amplitude $\Delta p_\mathrm{a}$ of the emitted acoustic waves is immaterial to the presented results. For all cases, the Lagrangian wave tracking is performed with a time-step of $\Delta t = 0.001 / f_\mathrm{a}$, which is sufficiently small such that the time-step does not influence the results.} The accuracy of the results obtained with the proposed model is assessed by comparison with results generated using the open-source software {\tt Wave-DNA} (v1.2), available at \url{https://doi.org/10.5281/zenodo.8229898}, which solves the convective Kuznetsov equation, \citep{Schenke2023} \changes{a second-order nonlinear wave equation that accounts for the background flow,} using a finite-dif\-ference method for domains with moving wave-emitting boundaries. \citep{Schenke2022}

\subsection{Acoustic black hole}
\label{sec:results_abh}

\begin{figure*}
    \centering
    \includegraphics[width=0.9\linewidth]{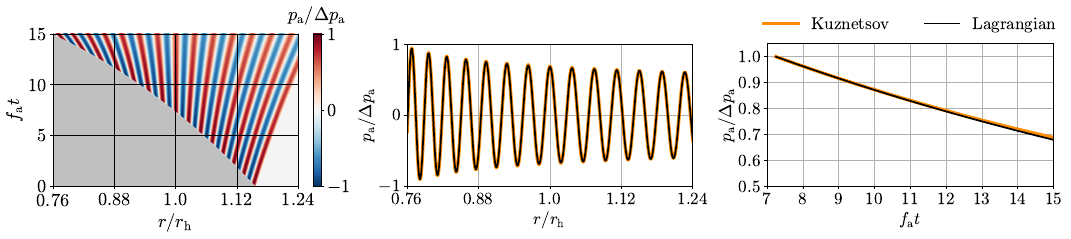}
    \caption{Space-time diagram (left), instantaneous wave profile at $f_\mathrm{a}t=15$ (center) and temporal evolution at the sonic horizon $r_\mathrm{h}$ (right) of the acoustic pressure $p_\mathrm{a}$ for a spherical acoustic black hole with $\mathrm{He}=40$, obtained using the proposed Lagrangian model. Results obtained by solving the convective Kuznetsov equation using a finite-difference method are shown for reference.}
    \label{fig:result_abh}
\end{figure*}

A spherical wave-emitting acoustic black hole is considered to validate the proposed correction of the Lagrangian wave tracking algorithm.
The radius of the wave-emitting boundary of the acoustic black hole is defined by \citep{Schenke2022d} $R(t) = (R_0^{3} - 3 c r_{\mathrm{h}}^2 t)^{\frac{1}{3}}$, where the initial radius $R_0$ of the emitter is chosen such that the pressure peak emitted at $7.25 \, f_\mathrm{a}t$ coincides with the sonic horizon. 
\changes{The Helmholtz number of this acoustic black hole is $\mathrm{He} = r_\mathrm{h}/\lambda_\text{a} = 40$.}
Fig.~\ref{fig:result_abh} shows the wave characteristics in the space-time plane, the instantaneous wave profile at $f_\mathrm{a}t = 15$, and the temporal evolution of the acoustic pressure $p_\mathrm{a}$ at the sonic horizon $r_\mathrm{h}$. The results obtained with the proposed model are in excellent agreement with the reference results obtained by solving the convective Kuznetsov equation, validating the correction proposed for the Lagrangian wave tracking algorithm.

\subsection{Pulsating spherical emitter}
\label{sec:results_emitter}

\changes{As a second case, a pulsating spherical emitter that induces a subsonic flow and simultaneously emits acoustic waves is considered, with the aim of testing the universal applicability of the presented quantification of the acoustic pressure modulation.} The time-varying radius of the emitter is defined as $R(t) = R_0 + \Delta \dot{R}_\mathrm{b} \, \sin (2\pi f_\mathrm{b} t)/(2 \pi f_\mathrm{b})$, where $f_\mathrm{b}=f_\text{a}/10$ is the frequency and $\Delta \dot{R}_\mathrm{b} = c/4$ is the velocity amplitude of the wave-emitting boundary. The flow induced by this boundary motion is subsonic, with a Mach number of $|u|/c \leq 0.25$. \changes{The initial wavelength $\lambda_\text{a}$ of the emitted acoustic waves follows from the Helmholtz number $\text{He}=R_0/\lambda_\text{a}$.}

\begin{figure*}
    \centering
    \includegraphics[width=0.9\linewidth]{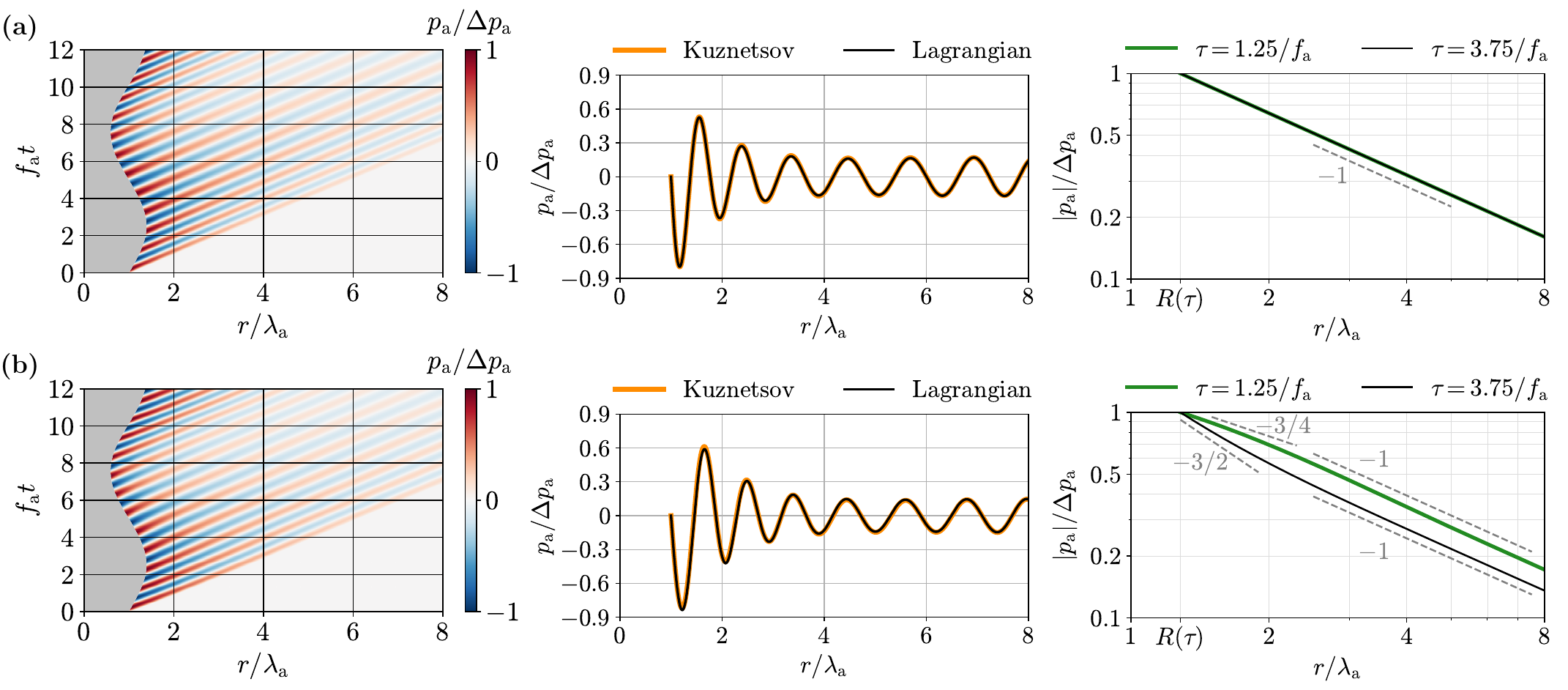}
    \caption{Space-time diagrams (left) and instantaneous acoustic wave profiles at $f_\mathrm{a}t=10$ (center) of the acoustic pressure $p_\mathrm{a}$, as well as $p_\mathrm{a}$ along two selected characteristics (right), emitted by a pulsating spherical emitter: (a) neglecting the induced flow and (b) taking the induced flow into account. Results obtained by solving the convective Kuznetsov equation using a finite-difference method are shown for reference.}
    \label{fig:result_emitters}
\end{figure*}

Fig.~\ref{fig:result_emitters} shows the acoustic pressure $p_\mathrm{a}$ in space-time diagrams, the instantaneous wave profile at $f_\mathrm{a}t = 10$, and along two representative characteristics, \changes{of a spherical emitter with $\text{He}=1$.
The results obtained by neglecting and accounting for the flow induced by the boundary motion of the emitter are shown in Fig.~\ref{fig:result_emitters}(a) and \ref{fig:result_emitters}(b), respectively. In both cases, the results obtained with the proposed Lagrangian model are in very close agreement with the reference results obtained by solving the convective Kuznetsov equation. However, comparing the wave profiles obtained by neglecting and accounting for the background flow, differences are observed close to the emitter,} which manifest clearly in the pressure evolution along individual characteristics. 

Neglecting the flow induced by the boundary motion, see Fig.~\ref{fig:result_emitters}(a), the acoustic pressure along  individual characteristics decays with $p_\mathrm{a} \propto 1/r$, \changes{as expected for a spherical acoustic wave in a quiescent fluid.} The profile of the emitted acoustic wave is more complex, since different parts of the wave are emitted at different emitter radii. Nevertheless, since $u = 0$, the modulation factor is $\mathcal{A} = 1$ and the acoustic pressure remains unmodulated. 

Taking the flow induced by the boundary motion into account, see Fig.~\ref{fig:result_emitters}(b), the acoustic pressure amplitude along individual characteristics is not proportional to $1/r$. The acoustic pressure emitted at $\tau = 1.25/f_\mathrm{a}$ propagates initially through \changes{an outward moving flow ($u > 0$)} driven by the expanding emitter ($\dot{R} > 0$) and, therefore, the spatial velocity derivative ${\partial u}/{\partial r} = - 2 {u}/{r}$  is negative. Hence, $\mathcal{A} > 1$ and the acoustic pressure decays slower than $1/r$, representing an effective increase of the acoustic pressure. In contrast, the acoustic pressure emitted at $\tau = 3.75/f_\mathrm{a}$ propagates initially through \changes{an inward moving flow ($u < 0$)} driven by the contracting emitter ($\dot{R} < 0$), with ${\partial u}/{\partial r}>0$. Hence, $\mathcal{A} < 1$ and the acoustic pressure decays quicker than $1/r$, representing an effective decrease of the acoustic pressure. \changes{This confirms the observations previously reported for acoustic black (white) hole analogues \citep{Schenke2023} that a positive (negative) spatial velocity derivative decreases (increases) the acoustic pressure.}

\begin{figure}
    \centering
    \includegraphics[width=0.3\textwidth]{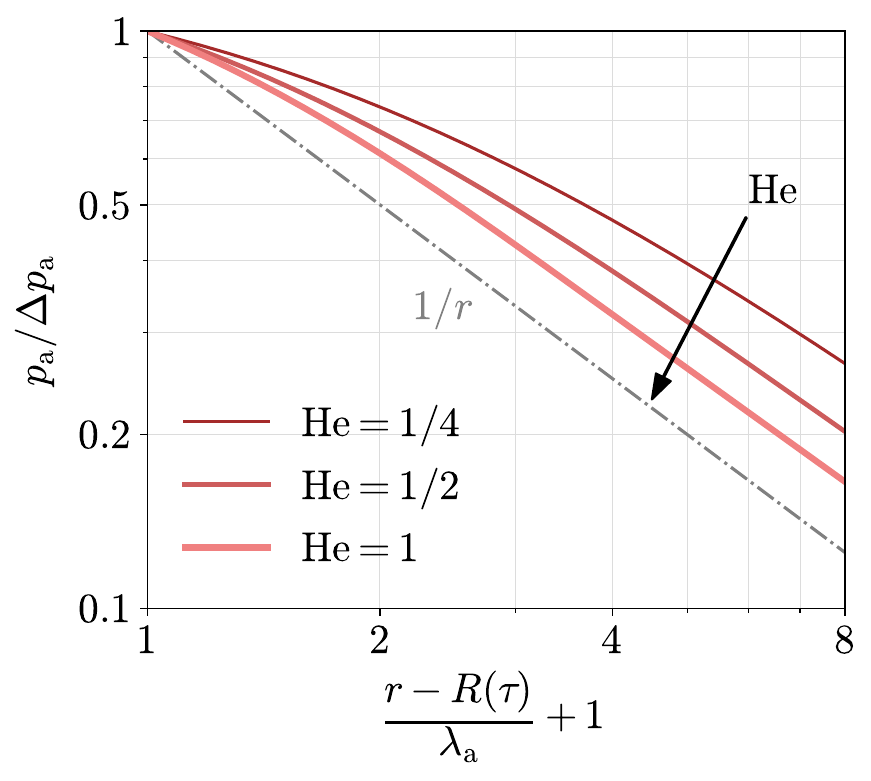}
    \caption{Acoustic pressure $p_\mathrm{a}$ along the characteristic emitted at $\tau = 1.25/f_\mathrm{a}$ by pulsating spherical emitters of different Helmholtz numbers $\mathrm{He}$, accounting for the induced flow.}
    \label{fig:result_emitters_scaling}
\end{figure}

As seen in Fig.~\ref{fig:result_emitters_scaling}, the acoustic pressure modulation is more pronounced for smaller Helmholtz numbers, $\mathrm{He} = {R_0}/{\lambda_\mathrm{a}}$, which corresponds to a more compact source with a spatially more rapidly changing flow.

\section{Conclusions}

A non-uniform flow changes the pressure of the acoustic waves propagating through it, a modulation that has recently been isolated and quantified using acoustic black/white hole analogues. \citep{Schenke2023}  
\changes{The presented results demonstrate that this pressure modulation is driven exclusively by spatially non-uniform flows. 
This cannot be inferred from the previously considered acoustic black or white hole analogues, \citep{Schenke2023} since the temporal derivative of the flow velocity is identically zero at the sonic horizon, $(\partial u/\partial t)_\mathrm{h} = 0$. However, being able to predict the acoustic pressure modulation driven by the spatially {and} temporally non-uniform flow field induced by the spherical emitter in Section \ref{sec:results_emitter}, using the modulation factor based only on the spatial velocity derivative proposed in Eq.~\eqref{eq:A}, implies that the temporal velocity derivative of the flow does not contribute to this modulation. Furthermore, the conservation of acoustic energy is sufficient to explain and quantify this pressure modulation, which suggests that any spatially non-uniform flow modulates the acoustic pressure.} 

\changes{The fact that the presented leading-order model of this acoustic pressure modulation can be readily applied to an acoustic black hole analogue as well as a subsonic non-uniform flow without changes supports the virtual acoustic black hole hypothesis and its claim that any acoustic wave characteristic forms locally an apparent sonic horizon. Consequently, observations made in acoustic black and white hole analogues may be generalized to arbitrary flows.}

To assess the influence of the observed acoustic pressure modulation in non-uniform three-dimensional flows, a next step may be to use numerical hydrodynamic/acoustic splitting methods employed in an Eulerian frame of reference, \citep{Ewert2021} where the flow is clearly separated from the acoustic perturbations and which, thus, allow to unambiguously isolate the identified pressure modulation and its influence.

\section*{Acknowledgements}
Fruitful discussions with Sören Schenke on acoustic black/white hole analogues are gratefully acknowledged.

\section*{Declaration of competing interest}
The author declares that he has no competing financial interests or personal relationships that could have appeared to influence the work reported in this paper.

\section*{Data availability}
The presented results have been generated using two open-source software tools, as indicated in the text, and can be reproduced by the readers with the provided information. 


\end{document}